\documentclass[preprint,12pt]{elsarticle}

\usepackage{amssymb}
\usepackage{amsthm}
\usepackage{amsmath}
\usepackage{bm}
\usepackage{xcolor}

\begin{document}

\begin{frontmatter}

\title{On-demand realization of topological states using Miura-folded metamaterials}

\author[inst1]{Shuaifeng Li}
\author[inst2]{Yubin Oh}
\author[inst2]{Seong Jae Choi}
\author[inst3]{Panayotis G.Kevrekidis}
\author[inst2]{Jinkyu Yang}

\affiliation[inst1]{organization={Department of Physics},
            addressline={University of Michigan}, 
            city={Ann Arbor},
            state={Michigan},
            postcode={48109},
            country={USA}}
\affiliation[inst2]{organization={Department of Mechanical Engineering},
            addressline={Seoul National University}, 
            city={Seoul},
            postcode={08826}, 
            country={Republic of Korea}}
            
\affiliation[inst3]{organization={Department of Mathematics and Statistics},
            addressline={University of Massachusetts}, 
            city={Amherst},
            state={Massachusetts},
            postcode={01003}, 
            country={USA}}

\begin{abstract}
Recent advancements in topological metamaterials have unveiled fruitful physics and numerous applications. Whereas initial efforts focus on achieving topologically protected edge states through principles of structural symmetry, the burgeoning field now aspires to customize topological states, tailoring their emergence and frequency. Here, our study presents the realization of topological phase transitions utilizing compliant mechanisms on the facets of Miura-folded metamaterials. This approach induces two opposite topological phases, leading to topological states at the interface. Moreover, we exploit the unique folding behavior of Miura-folded metamaterials to tune the frequency of topological states and dynamically toggle their presence. Our research not only paves the way for inducing topological phase transitions in Miura-folded structures but also enables the on-demand control of topological states, with promising applications in wave manipulation and vibration isolation.
\end{abstract}



\end{frontmatter}

\section{Introduction}
\label{introduction}
Topological metamaterials represent an emerging class of engineered structures, known for their exceptional guiding of wave propagation~\cite{ni2023topological}. These metamaterials draw inspiration from topology, a branch of mathematics concerned with the properties of space that remain invariant under continuous deformation. By strategically arranging the elements within the unit cell based on principles of symmetry, engineers can fabricate metamaterials that facilitate wave propagation along edges while exhibiting robustness against various imperfections. Spanning a diverse range of applications from electromagnetics~\cite{lu2014topological}, acoustics~\cite{zhang2018topological,xue2022topological}, and mechanics~\cite{xin2020topological,xue2022topological}, topological metamaterials pave the way for efficient guiding of waves and robust signal processing. Specifically, advancements of topological elastic metamaterials have led to materials capable of guiding stress waves and mitigating vibrations, thereby enhancing the structural robustness~\cite{xin2020topological}. Despite the considerable progress made in realizing topological states through various designs in recent years, a significant challenge lies in designing topological metamaterials with properties of on-demand topological states~\cite{bertoldi2017flexible,wang2020tunable}. Achieving such properties requires metamaterials capable of intrinsic adjustment of geometry and materials properties under the external stimuli, which is difficult for conventional metamaterials~\cite{li2019valley,li2021topological}. Soft metamaterials have emerged as a promising solution for achieving tunable topological states~\cite{li2018observation,niu2020reliable,niu2021interface}. The design allows for the modulation of topological states by exploiting the inherent deformability of soft materials—a strategy that utilizes material flexibility to adjust the emergence and frequency of topological states. However, the energy dissipation in such soft materials introduces practical difficulties for waveguide applications in large scales. Alternatively, exploring structural flexibility—a concept where hard materials are made transformable through mechanical mechanisms—might hold promising potential in this field.

Amid this landscape, origami-inspired engineering has revolutionized the field, diversifying from its ancient paper folding origins toward a sophisticated scientific discipline with wide-ranging applications. The essence of origami lies in its intricate designs emerging from a series of folds on a flat sheet of paper, resulting in different patterns and geometries~\cite{meloni2021engineering,zhu2022review}. In recent years, origami has been extended into the field of origami metamaterials, exhibiting remarkable mechanical properties, including tunable Poisson’s ratio~\cite{pratapa2019geometric,lyu2021origami}, exceptional rigidity~\cite{filipov2015origami}, and the ability to morph into multistable states~\cite{jianguo2015bistable,hanna2014waterbomb,yasuda2017origami}, which holds significant promise across many fields, from biomedical devices to aerospace engineering. Of particular interest is the ability of origami metamaterials to adapt their mechanical behaviors through fold manipulation.
This is an attribute that has inspired applications such as impact-absorbing systems~\cite{ji2021vibration,yasuda2019origami,zhou2016dynamic}, deployable structures that compactly fold for transport and expand on-site~\cite{chen2012folding,leanza2024active}, focusing and guiding of acoustic waves~\cite{harne2016origami,babaee2016reconfigurable,zhang2023tunable}. Furthermore, origami metamaterials have showcased potential in condensed matter physics, particularly in the field of topological metamaterials~\cite{miyazawa2022topological,li2023geometry,li2024emergence,li2024topological}. In light of the aforementioned challenges in the topological metamaterials, the tunability of origami metamaterials offers a plausible avenue for realizing on-demand topological states through the use of flexible structures.

In our study, we employ the Miura-folded metamaterials to realize a topological phase transition by integrating compliant mechanisms into the facets. These mechanisms induce opposite topological phases depending on whether they are integrated via inward or outward creases, resulting in the formation of topological states at the interfaces. Besides, the Miura-folded approach enables the tuning of the frequency of the topological states. Also, by leveraging the folding behavior, we can effectively switch topological states on and off, thus demonstrating versatile control over both frequency and presence of topological states. This capability to achieve on-demand topological states in Miura-folded metamaterials paves the way for manipulating elastic waves within flexible structures, such as origami-based systems, with potential applications in waveguides and vibration control.

\section{Methods\label{Design}}
\begin{figure}[h!]
    \centering
    \includegraphics[width=0.8\textwidth]{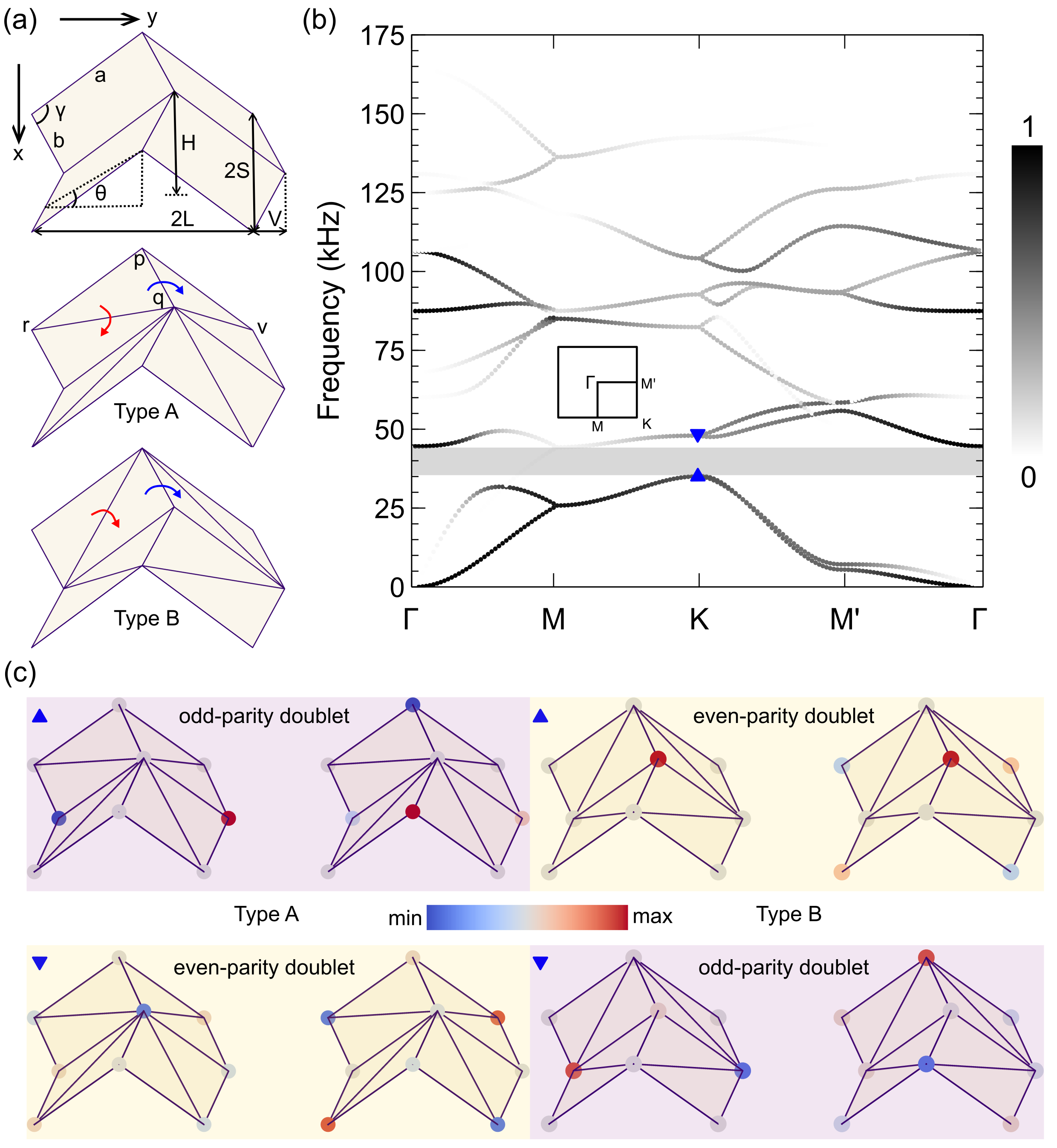}
    \caption{Topological phase transition by introducing compliant mechanisms on the facets.
    (a) From top to bottom, the unit cells of Miura-folded metamaterials without compliant mechanism on facets, inward compliant mechanism on facets and outward compliant mechanism on facets are shown, respectively. In the bar-and-hinge model, purple lines represent the bar elements with bar stiffness. The blue and red arrows represent the folding of the facets and bending of the facets, respectively.
    (b) The band structure of Miura-folded metamaterials with inward compliant mechanism on facets and outward compliant mechanism on facets. Note that the band structures for the two types of designs are identical. The polarization along the $z$ direction is encoded by the color, where $1$ represents the complete flexural modes and $0$ represents the complete in-plane modes.
    (c) The modes [$\mathrm{Re}\left(u_{z}\right)$] at $\bm{K}$ point in the Brillouin zone of the first four bands. The modes represented by the odd-parity doublet and even-parity doublet are inverted for the Miura-folded metamaterials with inward compliant mechanism and outward compliant mechanism.
    \label{fig:fig1}}
\end{figure}

In this section, we introduce the framework for structural analysis of Miura-folded metamaterials, laying the foundations for the subsequent eigen analysis, and time-dependent wave propagation. The Miura-folded metamaterials are featured by their ability to be formed from the flat state and to be folded with only bending along the folding creases, allowing the entire structure to be crafted from one sheet and compressed into a reduced volume. Within the structure of Miura-folded metamaterials, the shapes formed by the folding lines are known as facets, while the locations at which these folding lines converge are called vertices. This geometric arrangement is illustrated in the unit cell geometry of Miura-folded metamaterials, depicted in the top panel of Figure~\ref{fig:fig1}(a). Here, $a$ and $b$ denote the side lengths of quadrilateral facets and $\gamma\in[0,\frac{\pi}{2}]$ is the acute angle. The dimensions of the unit cell $H$, $S$, $L$ and $V$ are expressed as below~\cite{schenk2013geometry}:
\begin{equation}
    \label{equ:equ1}
    H=a\sin{\theta}\sin{\gamma},
\end{equation}
\begin{equation}
    \label{equ:equ2}
    S=b\frac{\cos{\theta}\tan{\gamma}}{\sqrt{1+\cos^2{\theta}\tan^2{\gamma}}},
\end{equation}
\begin{equation}
    \label{equ:equ3}
    L=a\sqrt{1-\sin^2{\theta}\sin^2{\gamma}},
\end{equation}
\begin{equation}
    \label{equ:equ4}
    V=b\frac{1}{\sqrt{1+\cos^2{\theta}\tan^2{\gamma}}},
\end{equation}
where $\theta\in[0,\frac{\pi}{2}]$ is the dihedral folding angle between the facets and the $xy$ plane. The Miura-folded metamaterials can transform from a flat state~($\theta=0$) to a fully folded state~($\theta=\frac{\pi}{2}$). To investigate the dynamic behaviors of this structure, rather than using the finite-element analysis with shell elements which provides detailed deformation, we employ the well-known bar-and-hinge model commonly utilized in origami research to model Miura-folded metamaterials with the trade-off between the accuracy and computational efficiency~\cite{filipov2015origami,yasuda2017origami,pratapa2018bloch,filipov2017bar,bhovad2021physical}.

In this framework, the entire origami structure is modeled as a series of bars linked via pin joints. The folding motion is modeled as a compliant mechanism using rotational hinges around these bars. In relation to the actual continuous structure, as shown in the first panel of Figure~\ref{fig:fig1}(a), the bars are akin to the crease lines that border each facet, while the nodes are equivalent to the vertices. The rotation along crease lines and the bending of the facets are represented by rotational hinges that are assigned different stiffness. Consequently, the structural deformation behaviors are determined by the extension of the bars and the relative rotations of the facets around the crease lines.

In the bar-and-hinge model, we have
\begin{equation}
    \label{equ:equ5}
    \begin{bmatrix}
        \bm{C}\\ \bm{J}_{b} \\ \bm{J}_{f}
    \end{bmatrix}
    \bm{u}=
    \begin{bmatrix}
        \bm{e}\\ \bm{\Phi}_{b} \\ \bm{\Phi}_{f}
    \end{bmatrix},
\end{equation}
where $\bm{u}$ denotes the nodal displacement, $\bm{e}$ is the bar extension, $\bm{C}$ is the compatibility matrix mapping $\bm{u}$ to $\bm{e}$, $\bm{J}_{b}$ and $\bm{J}_{f}$ are rotation compatibility matrices mapping $\bm{u}$ to $\bm{\Phi}_{b}$~(rotation around bending crease lines) and $\bm{\Phi}_{f}$~(rotation around folding crease lines), respectively. Therein, $\bm{C}$ can be easily constructed by considering the elongation of the bar due to the infinitesimal displacement of two vertices in the linear regime. Regarding the rotation compatibility matrices $\bm{J}_{b}$ and $\bm{J}_{f}$, we use the method developed by Liu and Paulino~\cite{liu2017nonlinear} and used in the modeling of Miura fold~\cite{bhovad2021physical}. As shown in the second panel of Figure~\ref{fig:fig1}(a), the geometry of the rotational spring element contains four vertices $\left(r,p,q,v\right)$, two triangles $\triangle{pqr}$, $\triangle{pqv}$ and the rotation angle $\phi_{pq}$. $\phi_{pq}$ can be calculated as:
\begin{equation}
    \label{equ:equ6}
    \phi_{pq}=\eta\arccos{\left(\frac{\bm{m}\cdot \bm{n}}{|\bm{m}||\bm{n}|}\right)} \quad \mathrm{modulo} \quad 2\pi,
\end{equation}
\begin{equation}
    \label{equ:equ7}
    \eta=\begin{cases}
            \mathrm{sign}\left(\bm{m}\cdot \bm{r}_{pv}\right), & \bm{m}\cdot \bm{r}_{pv} \neq 0; \\
            1, & \bm{m}\cdot \bm{r}_{pv}=0.
        \end{cases}
\end{equation}
Here, $\bm{m}$ and $\bm{n}$ are the current surface normal vector of the triangles $\triangle{pqr}$ and $\triangle{pqv}$, as expressed as $\bm{m}=\bm{r}_{rq}\times \bm{r}_{pq}$ and $\bm{n}=\bm{r}_{pq}\times \bm{r}_{pv}$ such that $\bm{r}_{rq}$ denotes the vector pointing from vertex $r$ to vertex $q$. After taking the derivative of the folding angle $\phi_{pq}$ with respect to the current position vector of vertex $p$, $\bm{r}_{p}$, we can obtain:
\begin{equation}
    \label{equ:equ8}
    \frac{\partial\phi_{pq}}{\partial\bm{r}_{p}}=
    \left(\frac{\bm{r}_{pv}\cdot \bm{r}_{pq}}{|\bm{r}_{pq}|}-1\right)\frac{\partial\phi_{pq}}{\partial\bm{r}_{v}}
    -\left(\frac{\bm{r}_{rq}\cdot \bm{r}_{pq}}{|\bm{r}_{pq}|}-1\right)\frac{\partial\phi_{pq}}{\partial\bm{r}_{r}},
\end{equation}
where
\begin{equation}
    \label{equ:equ9}
    \frac{\partial\phi_{pq}}{\partial\bm{r}_{r}}=
    \frac{|\bm{r}_{pq}|}{|\bm{m}|^{2}}\bm{m},
\end{equation}
\begin{equation}
    \label{equ:equ10}
    \frac{\partial\phi_{pq}}{\partial\bm{r}_{v}}=
    -\frac{|\bm{r}_{pq}|}{|\bm{n}|^{2}}\bm{n},
\end{equation}
and, $\bm{r}_{r}$ and $\bm{r}_{v}$ are the position vectors of vertex $r$ and $v$, respectively. In this way, the rotation compatibility matrix for the bending crease $\bm{J}_{b}$ and that for the folding crease $\bm{J}_{f}$ can be conveniently constructed.

Then, the stiffness matrix $\bm{K}$ can be written as:
\begin{equation}
    \label{equ:equ11}
    \begin{split}
    \bm{K}&=\begin{bmatrix}
        \bm{C}^{T} & \bm{J}_{b}^{T} & \bm{J}_{f}^{T}
    \end{bmatrix}
    \begin{bmatrix}
        \bm{K}_{e} & 0 & 0\\
        0 & \bm{K}_{b} & 0\\
        0 & 0 & \bm{K}_{f}\\
    \end{bmatrix}
    \begin{bmatrix}
        \bm{C}\\ \bm{J}_{b} \\ \bm{J}_{f}
    \end{bmatrix}\\   &=\bm{C}^T\bm{K}_{e}\bm{C}+\bm{J}_{b}^{T}\bm{K}_{b}\bm{J}_{b}+\bm{J}_{f}^{T}\bm{K}_{f}\bm{J}_{f}.
    \end{split}
\end{equation}
where $\bm{K}_{e}$, $\bm{K}_{b}$ and $\bm{K}_{f}$ represent the axial stiffness of the bars, rotational stiffness along the bending crease lines and rotational stiffness along the folding crease lines, respectively.

After the mass matrix $\bm{M}$ is constructed by distributing facet mass uniformly to four vertices, the generalized eigenvalue problem in the wave propagation can be expressed as:
\begin{equation}
    \label{equ:equ12}
    \bm{K}\bm{U}=\omega^2\bm{M}\bm{U},
\end{equation}
where $\omega$ is the angular frequency of harmonic wave propagation. To explore the wave propagation in the periodic structure, we use the unit cell analysis by using Bloch theorem $\bm{U}_{l,w}=e^{-i\bm{k}(la_{1}+wa_{2})}\bm{U}$ with basis vectors $a_{1}=[2S,0]$ and $a_{2}=[0,2L]$. $\bm{k}$ is the Bloch wave vector and $l$, $w$ denote the site of the unit cell. For the time-domain wave propagation in the Miura-folded metamaterials, we use the fourth order Runge-Kutta method to solve the governing equation of motion
\begin{equation}
    \label{equ:equ13}
    \bm{M}\Ddot{\bm{u}}+\bm{K}\bm{u}=\bm{F},
\end{equation}
under the excitation of the harmonic force $\bm{F}$.

\section{Results and Discussions}
\subsection{Topological phase transition}
In our Miura-folded metamaterials, we maintain fixed quadrilateral sides with dimensions $a=b=20~\mathrm{mm}$ and the acute angle $\gamma=80^{\circ}$. The folding angle is tunable and is chosen to be $\theta=40^{\circ}$ for demonstration in Figure~\ref{fig:fig1}. The bar stiffness, bending stiffness and folding stiffness are $1\times 10^{8}~\mathrm{N/m}$, $5\times 10^{2}~\mathrm{N\cdot{m}/rad}$ and $5\times 10^{1}~\mathrm{N\cdot{m}/rad}$, respectively, to simulate the Miura-folded metamaterials made of hard materials~\cite{filipov2017bar,schenk2011cold,xiang2021mechanical}. Since we choose $k_{b}=10 k_{f}$, the folding behavior is primarily governed by changes in dihedral angles around the folding crease lines instead of the bending crease lines. Note that without further notice, in the analysis throughout our entire work, the coordinate system follows what is shown in Figure~\ref{fig:fig1}(a), i.e., the downward and rightward directions are $x$ and $y$ directions, respectively. To create two opposite topological phases, we introduce inward bending creases on the facets~(type A) and outward bending creases on the facets~(type B), as depicted in the second and third panels of Figure~\ref{fig:fig1}(a). In the bar-and-hinge model, the compliant mechanisms on the facet serve as bars as well as the bending hinges of the facet.

Following the bar-and-hinge model illustrated in the previous section, we calculate the band structure $f(\bm{k})=\frac{\omega(\bm{k})}{2\pi}$ along $\Gamma$--$M$--$K$--$M'$--$\Gamma$ direction in the Brillouin zone shown in Figure~\ref{fig:fig1}(b). Note that band structures for type A and type B are identical. To identify the flexural wave mode, the polarization coefficient $\alpha=\frac{\sum_{i=1}^{i=4}|u_{zi}|^{2}}{\sum_{i=1}^{i=4}\left(|u_{xi}|^{2}+|u_{yi}|^{2}+|u_{zi}|^{2}\right)}$ is calculated and encoded by the gray-gradient color, with the modes of the first two bands predominantly exhibiting flexural modes and those of the third and fourth bands showing the weaker flexural modes. $\left(u_{x},u_{y},u_{z}\right)$ is the complex displacement field. We then compare the flexural modes~[the real part of eigen modes along $z$ direction $\mathrm{Re}\left(u_{z}\right)$] between the type A and type B structures in Figure~\ref{fig:fig1}(c). On the left panel, the odd-parity doublet comprising the $p_{x}$ and $p_{y}$ orbital--like states manifests in the first two bands, while the even-parity doublet consisting of the $s$ and $d$ orbital--like states appears in the third and fourth bands. Conversely, on the right panel, the even-parity doublet is evident in the first two bands, while the odd-parity doublet appears in the third and fourth bands. Note that the $p_{x}$~($p_{y}$) state is asymmetric with respect to the center, even symmetric to the $x$--~($y$--) axis, and odd symmetric to the $y$--~($x$--) axis, while the $s$ and $d$ states are symmetric with respect to the center. This observed band inversion implies the topological phase transition. Therefore, it is expected that the topological states will emerge within the bandgap region along the interface formed by Miura-folded metamaterials exhibiting these two topological phases.

\subsection{On-demand topological states in Miura-folded metamaterials}
\begin{figure}[h!]
    \centering
    \includegraphics[width=0.6\textwidth]{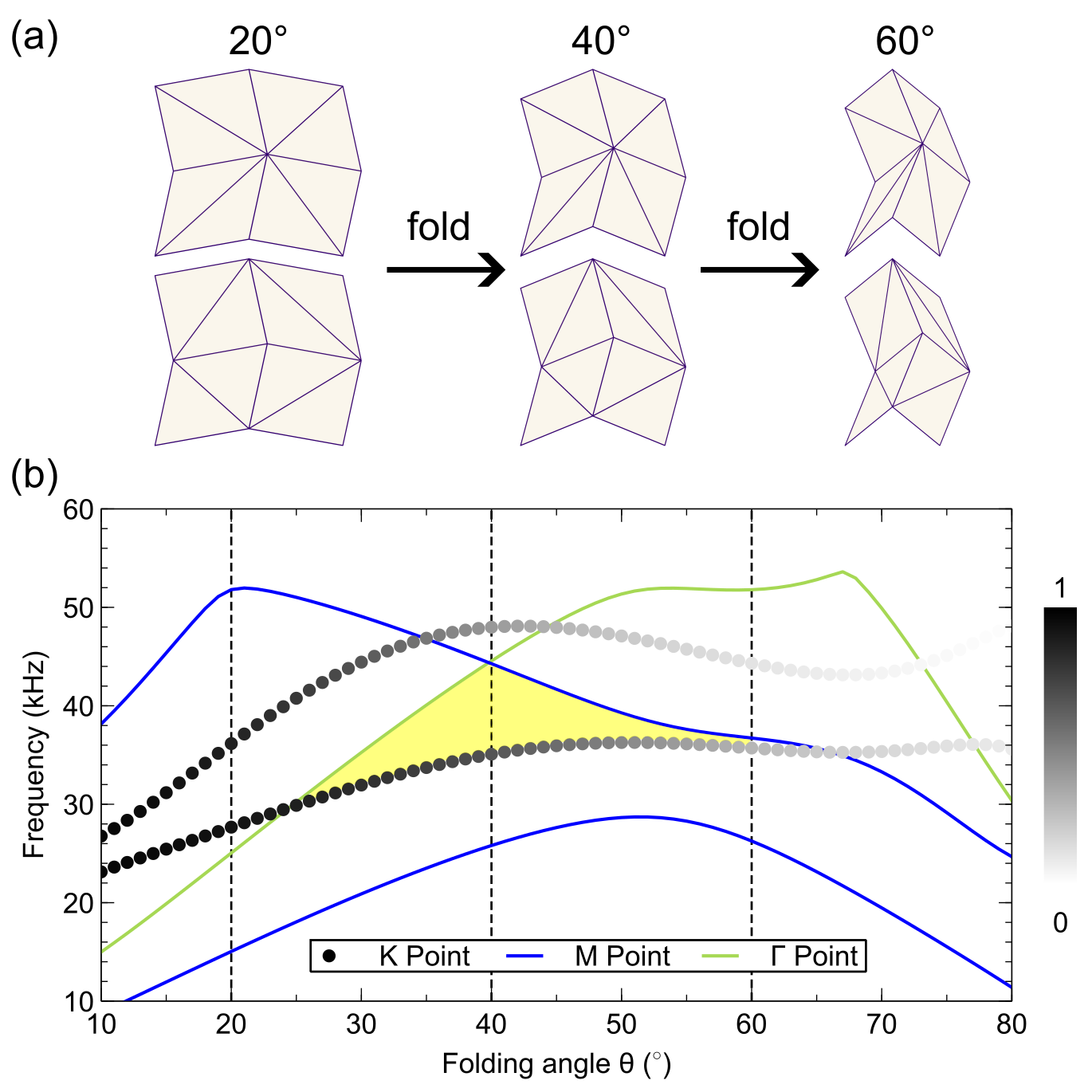}
    \caption{Tunable bandgap by folding Miura-folded metamaterials.
    (a) From left to right, the configurations of unit cells with folding angles $20^{\circ}$, $40^{\circ}$ and $60^{\circ}$ are shown, respectively.
    (b) The variation of the frequency of four bands at $K$, $M$ and $\Gamma$ points as a function of the folding angle $\theta$. The yellow shaded region represents the full bandgap for flexural waves. The gray-encoded dotted lines represent the polarization coefficient. The black dashed lines denote the folding angles corresponding to three cases in (a).
    \label{fig:fig2}}
\end{figure}

A key characteristic of Miura-folded metamaterials is the folding behaviors. Figure~\ref{fig:fig2}(a) illustrates the schematics of the unit cells for folding angles of $20^{\circ}$, $40^{\circ}$ and $60^{\circ}$. It is obvious that the change of the dimension along the $y$ direction~($L$) is much more than that along $x$ direction~($S$). The geometrical changes due to the folding angle $\theta$ result in evolution in the band structure. The intuitive expectation is that the shrinkage of the unit cell under folding can increase the frequency of the band gap. In Figure~\ref{fig:fig2}(b), we investigate the variation of frequency at high symmetry points, $\Gamma$, $M$ and $K$ points. Note that a large bandgap always exists along $\Gamma M'$ point as shown in Figure~\ref{fig:fig1}(b) and thus investigating the other three high symmetry points can provide a comprehensive view of bandgap evolution during folding. For small folding angles, a bandgap exists along $MK$ direction, but the bandgap along the $\Gamma M$ direction is eliminated due to the lower frequency at the $\Gamma$ point~(green line). As the folding angle $\theta$ increases, the frequency range of the bandgap along $MK$ and the frequency at the $\Gamma$ point increase, eventually resulting in the emergence of a full bandgap for the flexural modes, illustrated by the yellow shaded area in Figure~\ref{fig:fig2}(b). The largest bandgap appears at around $\theta=40^{\circ}$. However, with the further increase of $\theta$, the upper limit of the bandgap frequency decreases and the lower limit of the bandgap frequency slightly increases, resulting in the diminishing bandgap until closure occurs at around $\theta=60^{\circ}$. Although the full bandgap for flexural mode disappears, we should also notice that the bandgap along $KM'$ still exists, as well as the bandgap along $\Gamma M'$, due to the opening of $K$ and $M'$ points.

Furthermore, we explore the variation of flexural modes at $K$ point represented by the polarization coefficient $\alpha$ as a function of folding angle $\theta$. As shown in the color-encoded dotted lines in Figure~\ref{fig:fig2}(b), $\alpha$ decreases with the increase of $\theta$. It is understandable that the Miura-ori turns to be more resistant to the out-of-plane deformation when it gradually becomes the folded state. Hence, flexural modes become less dominant. Throughout the folding process, we can observe the frequency change of bandgap, the opening and closing of bandgap, and the variation of flexural modes, thereby presenting opportunities to achieve on-demand topological states.

\begin{figure}[h!]
    \centering
    \includegraphics[width=0.8\textwidth]{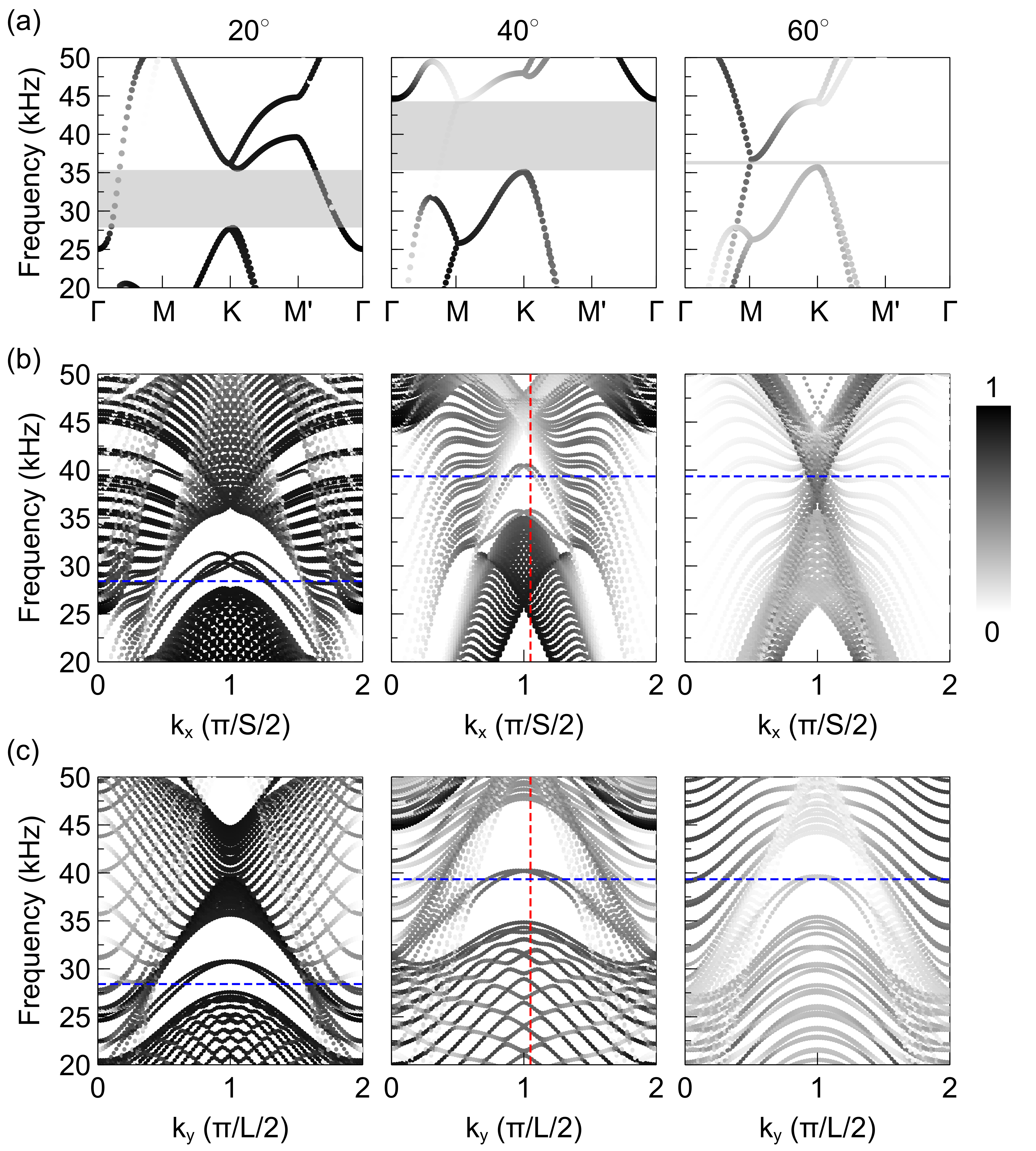}
    \caption{Tunable band structure and projected band structure by folding Miura-folded metamaterials.
    (a) the band structures, (b) the projected band structure along the $k_{x}$ direction and (c) the projected band structure along the $k_{y}$ direction when the folding angles are $20^{\circ}$, $40^{\circ}$ and $60^{\circ}$ are shown from left to right, respectively. The polarization coefficients are encoded by the color. The blue lines in (b) and (c) represent the excitation frequencies of the corresponding systems. The red lines in the second panel of (b) and (c) denote small offsets from the origin when $\delta k_{x}=\frac{\pi}{60S}$ and $\delta k_{y}=\frac{\pi}{60L}$, respectively, to analyze the eigen modes in the following.
    \label{fig:fig3}}
\end{figure}

We then show detailed band structures of these three folded states. As shown in Figure~\ref{fig:fig3}(a), from the left to right panel, the band structures for $\theta=20^{\circ}$, $\theta=40^{\circ}$ and $\theta=60^{\circ}$ demonstrate clear variations in the frequency of bandgap~($\theta=40^{\circ}\rightarrow \theta=20^{\circ}$), and the opening and closure of bandgap~($\theta=40^{\circ}\rightarrow \theta=60^{\circ}$). Besides, when $\theta=20^{\circ}$, the first four bands are dominated by flexural modes~(see the heavy gray dotted lines). However, with the increase of $\theta$, the flexural modes gradually become less dominant in the low-frequency bands~(see the light gray dotted lines when $\theta=40^{\circ}$ and $\theta=60^{\circ}$), which coincides with the investigation in Figure~\ref{fig:fig2}(b).

The projected band structures along $k_{x}$~[Figure~\ref{fig:fig3}(b)] and $k_{y}$~[Figure~\ref{fig:fig3}(c)] directions further elucidate these variations. The projected band structure is calculated using a ribbon structure A--B--A with $10$ unit cells for each type. The periodic boundary condition is applied along the $k_{x}$ direction or $k_{y}$ direction and the free boundary condition is applied to the two ends of the ribbon. When $\theta=20^{\circ}$ and $\theta=40^{\circ}$, the topological states emerge within the bandgap along the $k_{x}$ direction at around $k_{x}=\frac{\pi}{2S}$, localized at difference interfaces~(A--B interface and B--A interface, which will be identified later in the eigen mode analysis). However, when $\theta=60^{\circ}$, the topological states merge into the bulk states due to the closure of the band gap along the $MK$ direction.

Figure~\ref{fig:fig3}(c) displays the projected band structures along the $k_{y}$ direction. Apart from the frequency variation of topological states, in comparison with the scenario along the $k_{x}$ direction, here, due to the perpetual opening of the bandgap along the $KM'$ direction, topological states persist within the bandgap at around $k_{y}=\frac{\pi}{2L}$. Besides, the localization characteristics of topological states can be observed simultaneously at both interfaces~(A--B interface and B--A interface, which will be discussed later in the eigen mode analysis), which is different from the case along the $k_{x}$ direction.

Meanwhile, the polarization coefficient changes for the topological states in each case. When $\theta=20^{\circ}$, the bands within the bandgap for both $k_{x}$ and $k_{y}$ directions are rendered by heavy gray, implying the dominating flexural modes. However, this dominance gradually disappears with the increasing $\theta$~(e.g., $\theta=40^{\circ}$, $\theta=60^{\circ}$), evidenced by the topological states rendered by light gray.

\begin{figure}[h!]
    \centering
    \includegraphics[width=1\textwidth]{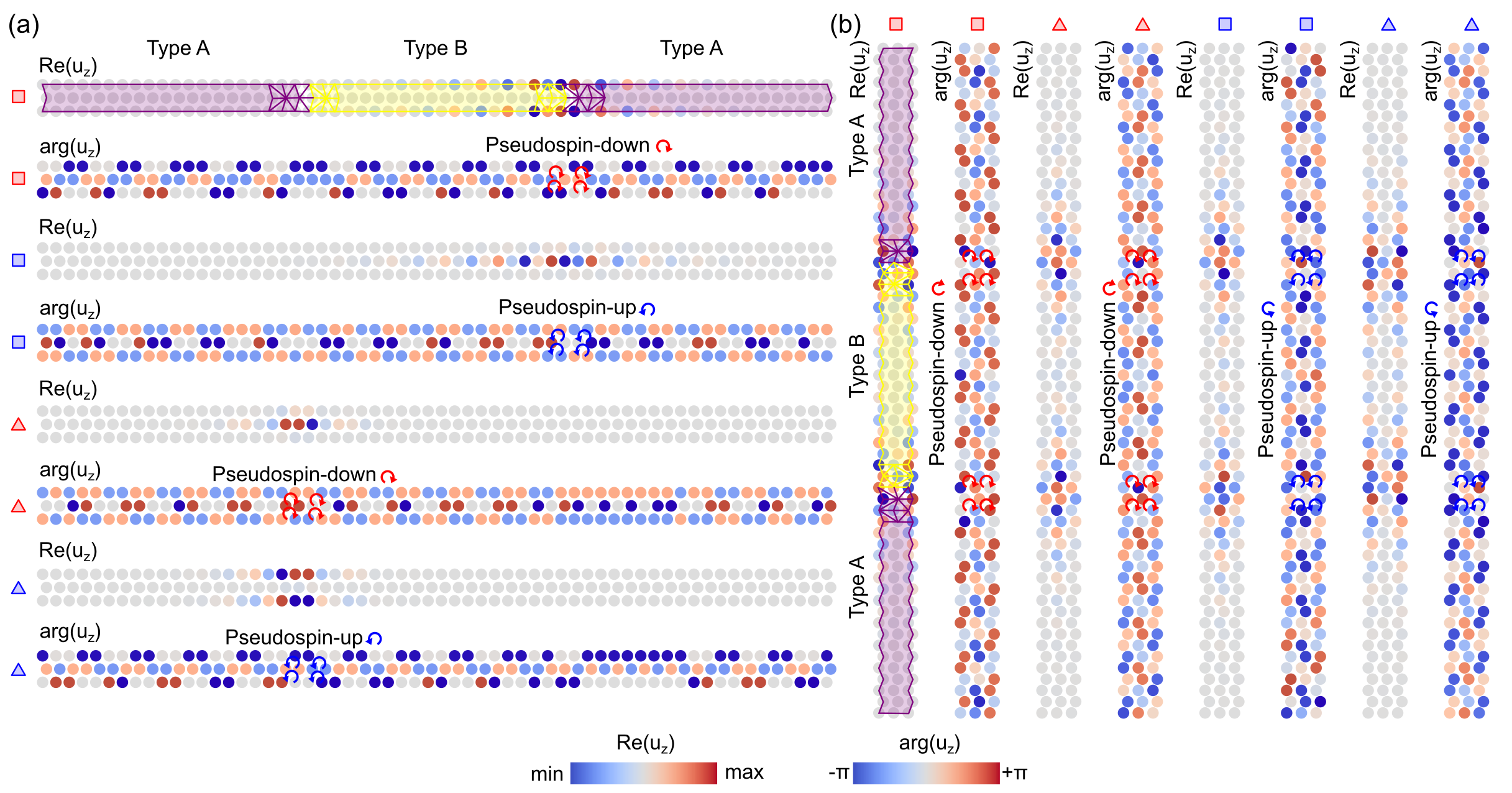}
    \caption{The real part and phase of the displacement field along $z$ direction corresponding to topological states when (a) $\delta k_{x}=\frac{\pi}{60S}$ and (b) $\delta k_{y}=\frac{\pi}{60L}$, where the localization of vertical displacement, the pseudospin-up and pseudospin-down states are observed. The real part and phase of displacement field along the $z$ direction are shown in sequence from top to bottom in (a) and from left to right in (b) to represent the topological states from low frequency and high frequency.
    \label{fig:fig4}}
\end{figure}

Next, we check the eigen modes~[$\mathrm{Re}(u_{z})$] and phases~[$\arg{ (u_{z})}$] of these topological bands appearing within the band gap to ensure the nature of the edge state and pseudospin in the analog of quantum spin Hall system~\cite{wu2023topological,liu2023second}. We take $\theta=40^{\circ}$ as an example and choose the $k_{x}=\frac{\pi}{2S}+\delta k_{x}$ for the states along the $k_{x}$ direction and $k_{y}=\frac{\pi}{2L}+\delta k_{y}$ for those along the $k_{y}$ direction, illustrated by the red dashed lines in the second panels of Figures~\ref{fig:fig3}(b) and \ref{fig:fig3}(c). As shown in Figure~\ref{fig:fig4}(a), from top to bottom, $\mathrm{Re}(u_{z})$ and the corresponding $\arg{(u_{z})}$ are shown from low frequency to high frequency when $\delta k_{x}=\frac{\pi}{60S}$. In general, the topological edge states can be observed by the localization of the displacement field. Specifically, the topological edge states manifest the interface-dependent property similar to topological valley systems. In the low frequency, the topological states are localized at the B--A interface, whereas they are localized at the A--B interface in the high frequency. We further investigate the pseudospin nature of the topological state by checking the phase~[$\arg{(u_{z})}$] of the vertices on the facets, where the pseudospin-up and pseudospin-down states can be observed.

In Figure~\ref{fig:fig4}(b), from left to right, $\mathrm{Re}(u_{z})$ and the corresponding $\arg{(u_{z})}$ are shown from low frequency to high frequency when $\delta k_{y}=\frac{\pi}{60L}$. The localization of the displacement field can be seen at both A--B and B--A interfaces simultaneously, distinct from the interface-dependent topological states along the $k_{x}$ direction. Besides, the pseudospin-up and pseudospin-down states can also be identified by the phases~[$\arg{(u_{z})}$] of vertices on the facets.

\begin{figure}[h!]
    \centering
    \includegraphics[width=0.8\textwidth]{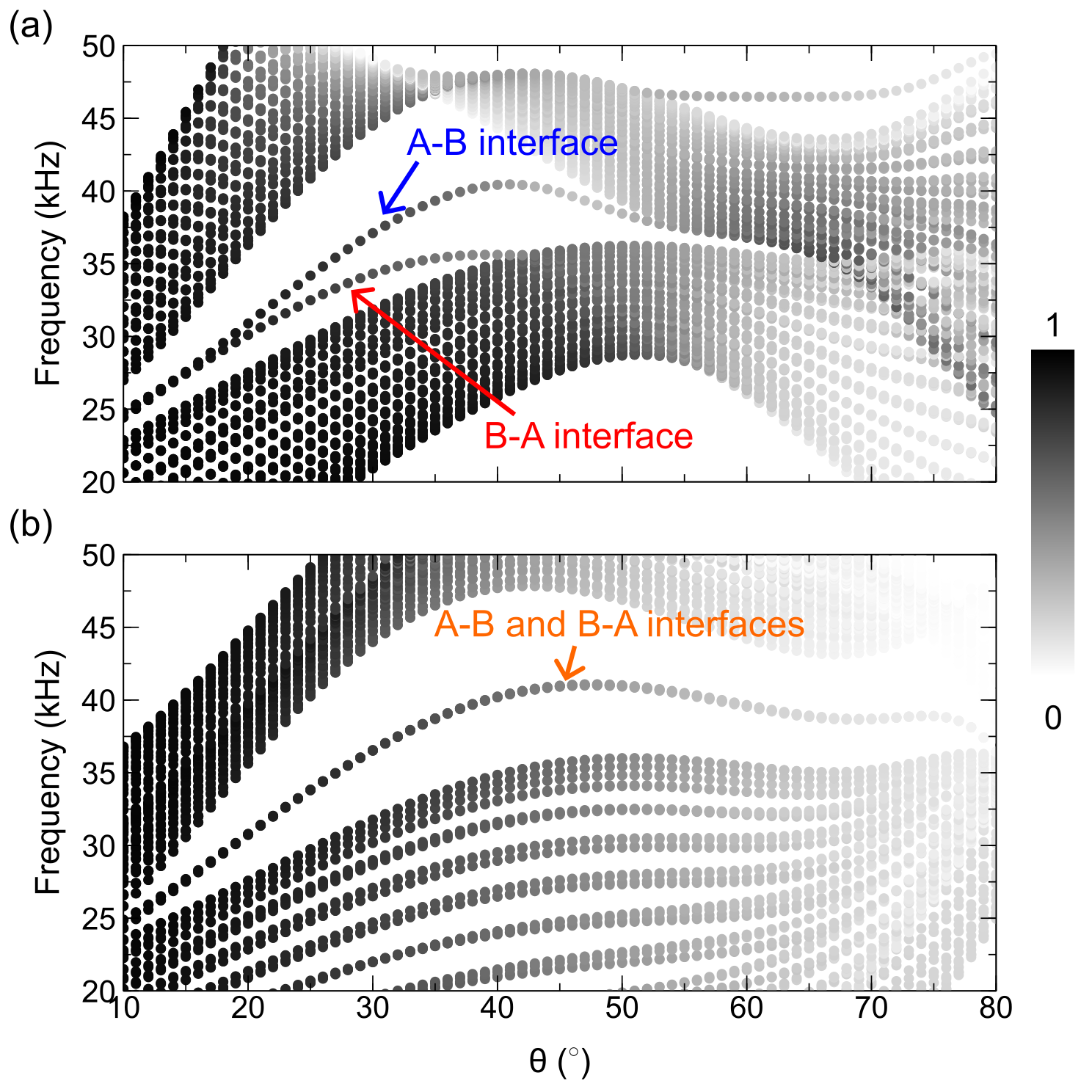}
    \caption{The variation of frequency in projected band structures when (a) $k_{x}=\frac{\pi}{2S}$ and (b) $k_{y}=\frac{\pi}{2L}$ by folding Miura-folded metamaterials from $\theta=10^{\circ}$ to $\theta=80^{\circ}$.
    \label{fig:fig5}}
\end{figure}

After the characterization of eigen modes, to thoroughly explore the evolution of topological states and bandgap along both the $k_{x}$ and $k_{y}$ directions by folding Miura-folded metamaterials, we calculate the projected band structure along $k_{x}$ under the fixed $k_{x}=\frac{\pi}{2S}$ and that along $k_{y}$ under the fixed $k_{y}=\frac{\pi}{2L}$ as a function of folding angle $\theta$. As shown in Figure~\ref{fig:fig5}(a), the frequencies of the bandgap and topological states increase as a function of $\theta$. After reaching the peak at around $\theta=40^{\circ}$, the topological states localized at B--A interface merge into the bulk states and the frequency of the topological states localized at A--B interface starts to decrease. Eventually, with the closure of the bandgap, topological states disappear.

The variation of projected band structure along $k_{y}$ direction~[Figure~\ref{fig:fig5}(b)] shows a similar trend to that along the $k_{x}$ direction~[Figure~\ref{fig:fig5}(a)]. As the folding angle increases, the frequency of the topological states increases. The frequency reaches the maximum at around $\theta=40^{\circ}$ and slightly decreases when $\theta$ continues to increase. Different from the case along the $k_{x}$ direction, the topological states along the $k_{y}$ direction always exist due to the perpetual opening of the bandgap along the $KM'$ direction.

Apart from the change on the frequency over folding, the polarization coefficients of the topological states also experience variation. Generally, with the increase of $\theta$, the flexural modes gradually become weaker in the topological states. The intriguing evolution of topological states on the frequency, emergence and modes along different directions opens up an avenue to manipulate elastic waves in our platform.

\subsection{Wave propagation in Miura-folded metamaterials}
\begin{figure}[h!]
    \centering
    \includegraphics[width=1\textwidth]{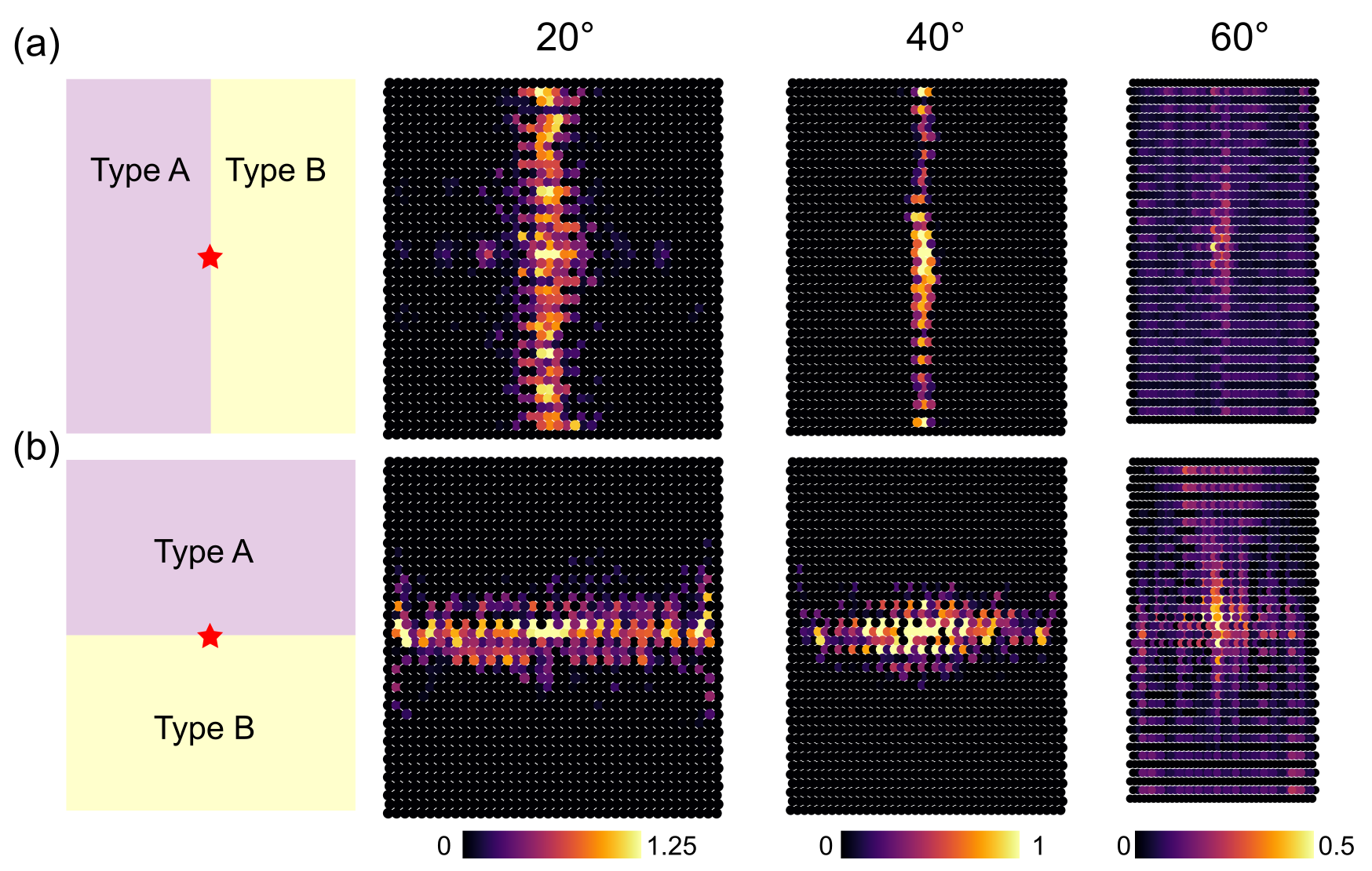}
    \caption{Wave propagation along the straight interface in Miura-folded metamaterials.
    The configurations for wave propagation along $k_{x}$ and $k_{y}$ directions are shown in the first panels in (a) and (b), respectively. The red stars represent the excitation position. From the second panel to the fourth panel, the RMS of $v_{z}$ of (a) the straight interface along $x$ direction, of (b) the straight interface along $y$ direction when the folding angles are $20^{\circ}$, $40^{\circ}$ and $60^{\circ}$ are shown, respectively.
    \label{fig:fig6}}
\end{figure}

After the eigen analysis above, we  conduct time-domain simulations to explore wave propagation in Miura-folded metamaterials under different folding angles. We start from the straight interface along both $k_{x}$ and $k_{y}$ directions by applying a harmonic force $F=F_{0}\cos(2\pi ft)$ along the $z$ direction at the middle of the interface, whose configurations are composed of $20\times 20$ unit cells shown in the first panels of Figure~\ref{fig:fig6}. $F_{0}=1~\mathrm{N}$ and $t$ spans from $0~\mathrm{ms}$ to $0.8~\mathrm{ms}$. The fixed boundary condition is applied to four sides of the metamaterials. In Figure~\ref{fig:fig6}, we show the root mean square~(RMS) of $v_{z}$ at the folding angle of $\theta=20^{\circ}$, $\theta=40^{\circ}$ and $\theta=60^{\circ}$, respectively. The second column panels of Figure~\ref{fig:fig6} show the RMS at the frequency of $28415~\mathrm{Hz}$~[blue dashed lines in Figures~\ref{fig:fig3}(b) and (c)] along the straight interface $k_{x}$ and $k_{y}$ direction, respectively. It is clear that the vibration is localized along the interface. When the folding angle increases to $\theta=40^{\circ}$, the frequency is raised to $39340~\mathrm{Hz}$~[blue dashed lines in Figures~\ref{fig:fig3}(b) and (c)] to excite the topological states along the $k_{x}$ and $k_{y}$ straight interfaces, as displayed in the third column panels of Figure~\ref{fig:fig6}, respectively. Both show excellent localization of vibration around the interface.

When the Miura-folded metamaterial is folded to $\theta=60^{\circ}$, as indicated by the projected band structure~(Figure~\ref{fig:fig3}), the topological states along $k_{x}$ vanish, while the frequency of the topological states along $k_{y}$ barely changes. Hence, we choose the same excitation frequency as the previous case, $39340~\mathrm{Hz}$~[blue dashed lines in Figures~\ref{fig:fig3}(b) and (c)]. In the fourth panel of Figure~\ref{fig:fig6}(a), the disappearance of topological states is evident, with bulk states being excited instead. In the fourth panel of Figure~\ref{fig:fig6}(b), the topological states along the $k_{y}$ direction still persist to some extent, albeit with the harmonic force also inevitably exciting the bulk state in the $k_{x}$ direction, leading to the less localized displacement field distribution.

Apart from the tunable frequency and emergence of the topological states, the polarization variation analyzed in the previous section also affects the wave propagation. Note that we keep the amplitude of the excitation force the same $F_{0}=1~\mathrm{N}$ for the time-domain simulation and normalize the RMS of $v_{z}$ for each case using the case of $40^{\circ}$ as the reference. However, the RMS of $v_{z}$ decreases with the increase of the folding angle~($20^{\circ}\rightarrow40^{\circ}\rightarrow60^{\circ}$, see the scale of the colorbar). This is related to the declining flexural polarization with the increase of the folding angle.

\begin{figure}[h!]
    \centering
    \includegraphics[width=1\textwidth]{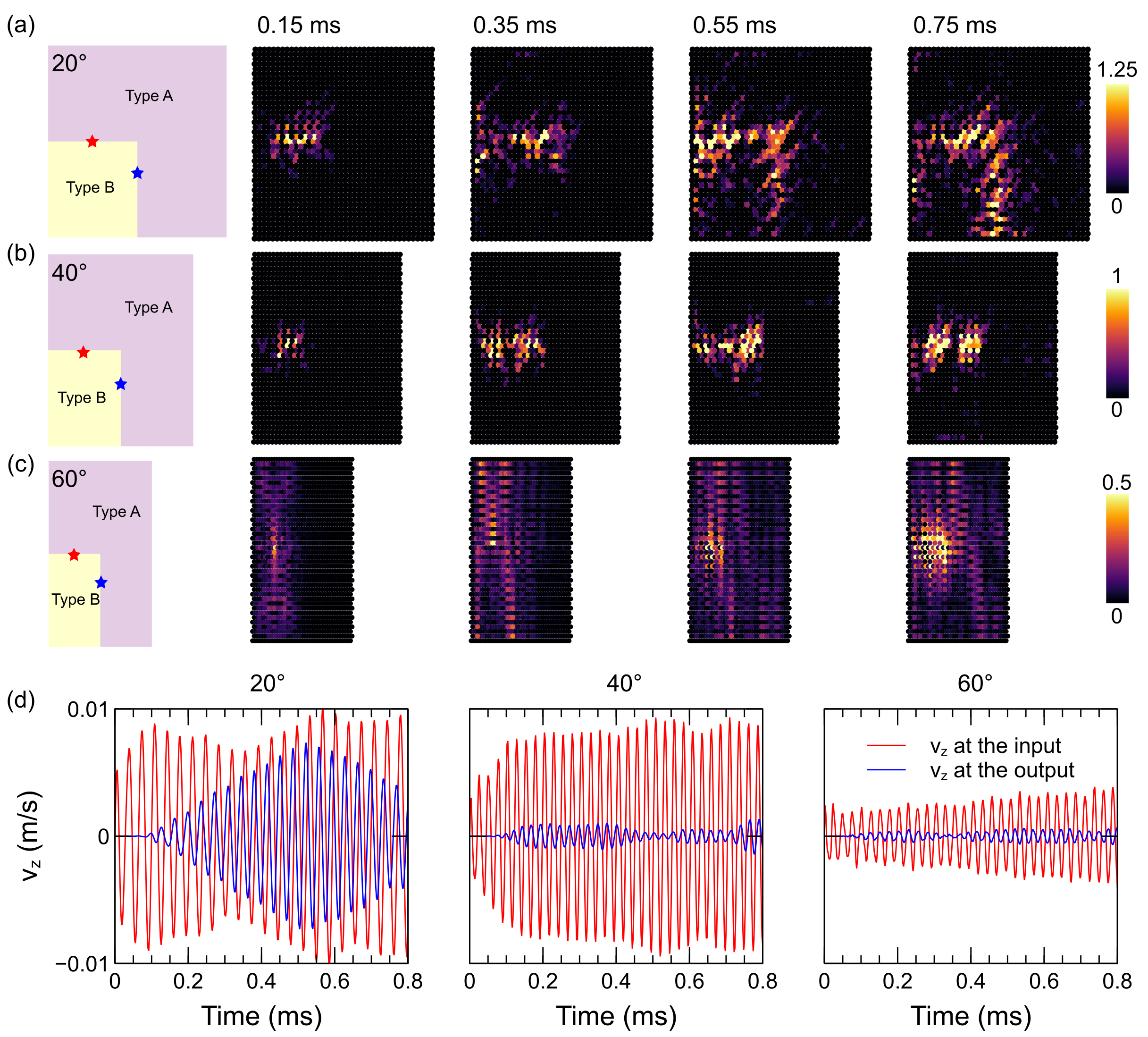}
    \caption{Wave propagation along the $L$-shaped interface in Miura-folded metamaterials.
    The configurations for wave propagation along the $L$-shaped interface are shown in the first panels in (a), (b) and (c), respectively. The red stars represent the excitation position. The amplitudes of $v_{z}$ in $0.15~\mathrm{ms}$, $0.35~\mathrm{ms}$, $0.55~\mathrm{ms}$ and $0.75~\mathrm{ms}$ are shown from the second panel to the fifth panel, respectively, for the folding angles (a) $20^{\circ}$, (b) $40^{\circ}$ and (c) $60^{\circ}$.
    (d) From left to right, $v_{z}$ at the input marked by red stars and $v_{z}$ at the output marked by blue stars for $20^{\circ}$, $40^{\circ}$ and $60^{\circ}$ are presented.
    \label{fig:fig7}}
\end{figure}

We also conduct time-domain simulations to explore wave propagation across the $L$-shaped interface. A harmonic force $F=F_{0}\cos(2\pi ft)$ along $z$ direction is applied at the interface, where $F_{0}=1~\mathrm{N}$. As depicted in the first panels of Figure~\ref{fig:fig7}, the inner structure is composed of the type B Miura-folded metamaterials while the outer structure contains type A Miura-folded metamaterials. The inner structure contains $10\times 10$ unit cells and the entire structure is composed of $20\times 20$ unit cells. The fixed boundary condition is applied to the four sides of the metamaterials. The interfaces associated with the wave propagation are A--B interface along $k_{y}$ direction and B--A interface along $k_{x}$ direction. As displayed in Figure~\ref{fig:fig7}(a), when the folding angle $\theta=20^{\circ}$, the excited waves propagate along the $L$-shaped interface with negligible backscattering at the bend. It indicates the excellent coupling between topological states at the A--B interface along the $k_{y}$ direction and those at the B--A interface along the $k_{x}$ direction due to the same frequency~[Figures~\ref{fig:fig3}(b) and (c)]. Meanwhile, we notice that waves are well localized along the interface with few leakages to the bulk.

In contrast, when the folding angle increases to $\theta=40^{\circ}$, as shown in Figure~\ref{fig:fig7}(b), the excited waves can propagate along the interface over time but experience reflection at the bend. This phenomenon manifests as reflected waves at around $0.55~\mathrm{ms}$, preventing the waves from traveling through the bend and reaching the output port. It indicates that the initial excited topological states at the A--B interface along $k_{y}$ cannot couple with those at the B--A interface along the $k_{x}$ direction, because the topological states at A--B interface along the $k_{y}$ direction are at the higher frequency while those at the B--A interface along the $k_{x}$ direction are at the lower frequency~[see Figures~\ref{fig:fig3}(b) and (c)].

At the same excitation frequency as the case of $\theta=40^{\circ}$ but with a folding angle $\theta=60^{\circ}$, as depicted in Figure~\ref{fig:fig7}(c), the waves propagate both along the interface due to the topological states along the $k_{y}$ direction and into the bulk due to the gap closing along the $k_{x}$ direction over time. In addition, despite the presence of topological states along the $k_{y}$ direction~[the third panel of Figure~\ref{fig:fig3}(c)], the localization of waves along the $k_{y}$ interface is not pronounced, overshadowed by the excitation of bulk states. Besides, similar to what is shown in Figure~\ref{fig:fig6}, because of the variation of the polarization, the amplitude of the excited flexural modes deceases as the folding angle increases.

In Figure~\ref{fig:fig7}(d), we show $v_{z}$ at the input~(red stars) and at the output~(blue stars) over time. When $\theta=20^{\circ}$, $v_{z}$ at the output and $v_{z}$ at the input are roughly in the same order of magnitude, indicating the negligible backscattering at the bend. However, when $\theta=40^{\circ}$ and $\theta=60^{\circ}$, $v_{z}$ at the output is significantly smaller than that at the input, demonstrating the stoppage of waves through the bend. Besides, the amplitude of $v_{z}$ at the input when $\theta=60^{\circ}$ is significantly smaller than that in other cases, indicating the decreased flexural modes, as analyzed in the previous section. Through the analysis of wave propagation in Miura-folded metamaterials at different folded states along both straight and bent interfaces, we showcase the tunability and switchability of topological states using Miura-folded metamaterials.

\section{Conclusions \& Future Challenges\label{conclusion}}
In conclusion, we introduce an approach to realizing topological phase transitions in Miura-folded metamaterials via the application of two distinct compliant mechanisms on the facets. This methodology results in the emergence of topological states along the interfaces formed by metamaterials with inward and outward compliant mechanisms. Besides, by harnessing the folding behaviors inherent in Miura-folded metamaterials, we demonstrate the realization of on-demand topological states, where the frequency of these states can be tuned~(e.g., $\theta=40^{\circ}\rightarrow \theta=20^{\circ}$) and their existence can be switched~(e.g., $\theta=40^{\circ}\rightarrow \theta=60^{\circ}$) by varying the folding angle. Moreover, by using the interface-dependent property of topological states along the $k_{x}$ direction, the wave propagation along the bend can be switched on and off as desired. In addition, the folding behaviors of Miura-folded metamaterials lead to the variation of flexural modes, which essentially manipulates the amplitude of the output.

Our study not only presents a method, which is experimentally feasible during the fabrication process of origami~\cite{miyazawa2023design}, to realize a topological phase transition by engineering the compliant mechanisms, but also shows the potential of flexible structures such as origami to serve as promising platforms for manipulating topological states and controlling wave propagation as desired. Indeed, one can envision different types of two-dimensional configurations, as well as different types of interfaces, leading to desired types of associated wave propagation, occurring at particular frequencies of interest. These findings have significant applications in tunable waveguides and vibration control, offering promising avenues for further research in this field.

J.Y. is grateful for the support by the New Faculty Startup Fund from Seoul National University. J.Y. also acknowledges the support from SNU-IAMD, SNU-IOER, and National Research Foundation grants funded by the Korea government [Grants No. 2023R1A2C2003705 and No. 2022H1D3A2A03096579 (Brain Pool Plus by the Ministry of Science and ICT)]. This material is based upon work supported by the U.S. National Science Foundation under the awards PHY-2110030 and DMS-2204702 (P.G.K.).

\bibliographystyle{elsarticle-num} 
\bibliography{references}

\end{document}